\documentclass[preprint2]{aastex}
\usepackage{graphicx}
\usepackage[small]{caption}
\usepackage{color}
\usepackage[all]{xy}

\usepackage{amsmath}
\usepackage{natbib}
\usepackage{epsfig}
\usepackage{txfonts}

\def\california{{NGC 1499 }}
\def\perseo{{G159.6-18.5 }}
\newcommand{\ions}[2]{{#1}\,{\sc #2}}

\begin{document}

\title{Constraints on the Polarization of the Anomalous Microwave Emission in
  the Perseus Molecular Complex from 7-year WMAP data}

\author{C.~H.~L\'opez-Caraballo, J.~A.~Rubi\~{n}o-Mart\'{i}n, R.~Rebolo,
  R.~G\'enova-Santos}
\email{clopez@iac.es, jalberto@iac.es, rrl@iac.es, rgs@iac.es }
\affil{Instituto de Astrof\'{i}sica de Canarias (IAC), C/V\'{i}a L\'{a}ctea,
  s/n, E-38200, La Laguna, Tenerife (Spain)}


\begin{abstract}
  We have used the seven year {\it Wilkinson Microwave Anisotropy Probe} (WMAP)
  data in order to update the measurements of the intensity signal in the
  \perseo region within the Perseus Molecular Complex, and to set constraints on
  the polarization level of the anomalous microwave emission in the frequency
  range where this emission is dominant.
  At 23, 33 and 41~GHz, we obtain upper limits on the fractional linear
  polarization of 1.0, 1.8 and 2.7\% respectively (with a 95 per cent confidence
  level).
  These measurements rule out a significant number of models based on magnetic
  dipole emission of grains that consist of a simple domain \citep{DraineMD} as
  responsible of the anomalous emission.
  When combining our results with the measurement obtained with the COSMOSOMAS
  experiment at 11~GHz \citep{Battistelli2006}, we find consistency with the
  predictions of the electric dipole and resonance relaxation theory
  \citep{Lazarian2000} at this frequency range.
\end{abstract}


\maketitle

\section{Introduction}
\label{sec:intro}

The dust-correlated microwave (10-60~GHz) emission detected by several cosmic
microwave background (CMB) experiments performed in the two last decades (COBE
\citep{Kogut1996}; OVRO at 14.5 and 32~GHz \citep{Leitch1997}; Saskatoon
\citep{Oliveira1997}; 19~GHz \citep{Oliveira1998}; Tenerife
\citep{Oliveira1999,Oliveira2004}; COSMOSOMAS
\citep{Watson2005,Hildebrandt2007}; VSA \citep{Scaife2007, Tibbs2010}; and
references therein) suggests the existence of a new continuum microwave emission
mechanism unlike to the three well-known Galactic mechanisms: synchrotron,
free-free and thermal dust emission.
A remarkable observational effort has been devoted to the understanding of the
intensity and polarization properties of this ``anomalous'' microwave emission,
among other reasons, because of the importance of an accurate foreground
correction of the CMB maps at low frequencies.

Among the various scenarios proposed to explain this emission, electric dipole
radiation \citep{DraineED} from very small (N $\leq 10^3$ atoms) rapidly
rotating ($\sim 1.5 \times 10^{10} \ {\rm s}^{-1}$) carbon based molecules in
the interstellar medium (the so-called ``{\it spinning dust}'') appears to
reproduce well the observational constraints \citep[see e.g.][]{Watson2005,
  Casassus2006, Iglesias2006, Dickinson2007, Tibbs2010}.
The detailed theoretical description of this family of models has been recently
updated in \citet{AliHaimoud2009}, also including the effect of rotation of the
dust grains around a non-principal axis \citep{Silsbee2010}.

An alternative explanation based on magnetic dipole emission of spinning dust
has also been proposed \citep{DraineMD}. Measurements of the polarization
properties of the anomalous microwave emission may potentially distinguish
between these two models. According to \cite{Lazarian2000}, electric dipole
radiation from spinning dust would be polarized at low frequencies, reaching a
maximum (6-7\%) at 2-3~GHz and dropping to 4-5\% at 10~GHz and progressively
decreasing at higher frequencies. Polarization from magnetic dipole emission
predicts a different frequency behaviour and stronger linear polarization
depending on the composition and shape of the emitting particles.

%
%
From the observational point of view, there is little information in the
literature about the polarization properties of the anomalous emission.
\cite{Kogut2007} used the full-sky WMAP 3-year data to constrain the
polarization fraction of a diffuse anomalous component traced by the dust
morphology. They concluded that the polarized anomalous emission contributes
less than 1\% of the observed polarization signal variance in any of the five
WMAP bands.

Only few attempts have been made to determine the polarization of the anomalous
microwave emission in individual objects, and in the frequency range where this
contribution is dominant (10-50~GHz). \cite{Battistelli2006}, using data from
the COSMOSOMAS experiment on the Perseus molecular complex, reported at 11~GHz
$\Pi = 3.4^{+1.5}_{-1.9} \%$ (95\% confidence level). More recently
\cite{Mason2009}, using the Green Bank Telescope at 9~GHz, obtained an upper
limit of linear polarization of 88~$\mu$K at 95.4\% of confidence in the Lynds
1622 dark cloud. Both the Perseus molecular complex and the Lynds 1622 cloud are
regions where anomalous microwave emission appears to dominate on other emission
processes in the frequency range 10-50~GHz.

In this work, we present new measurements of the polarization of the anomalous
microwave emission in \perseo (within the Perseus molecular complex) using the
WMAP ({\it Wilkinson Microwave Anisotropy Probe}) 7-year data. Although the
results only provide upper limits to the polarized emission in the region, they
still constitute a strong constraint on the physical mechanism responsible for
the emission.
%

\section{The G159.6-18.5 region}
\label{sec:perseusregion}

\begin{figure}
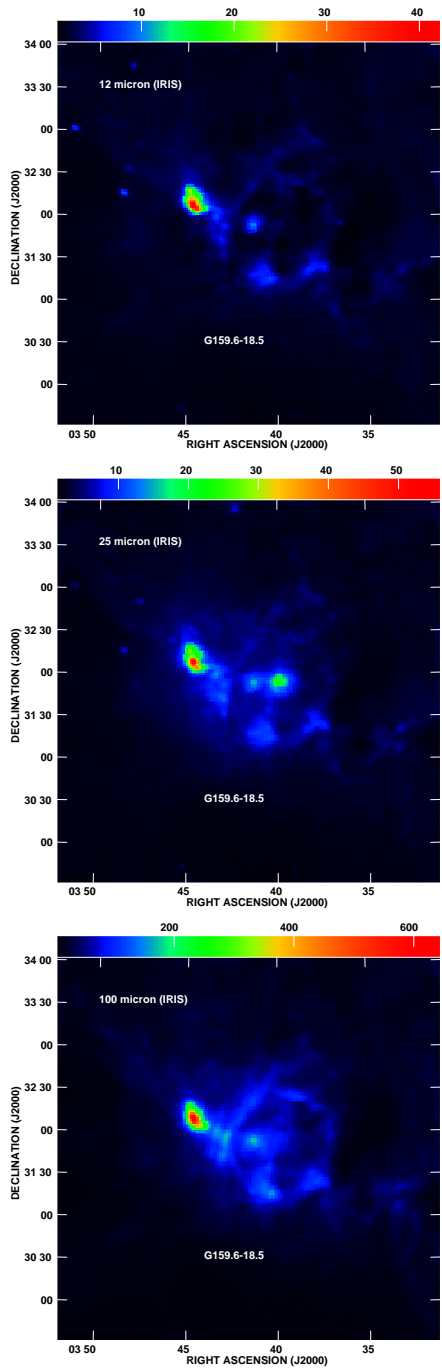

\centering
\includegraphics[width=6.0cm]{Perseo_iris12.ps}\\
\includegraphics[width=6.0cm]{Perseo_iris25.ps}\\
\includegraphics[width=6.0cm]{Perseo_iris100.ps}%
\caption{IRIS images at $12 \mu m$, $60 \mu m$ (middle) and $100 \mu m$ (bottom
  panel) of the \perseo region, with a field of view of $4.5^{\circ} \times
  4.5^{\circ}$, and centered in R.A.(J2000)$=55.4^\circ$ and
  Dec.$=+31.8^\circ$. All maps are in units of MJy~sr$^{-1}$. }
\label{ima:iras100}
\end{figure}

The Perseus molecular complex is a giant molecular cloud located at a distance
of 260~pc \citep{Cernicharo1985}.
Our region of interest is \perseo, a dust feature in this molecular complex,
observed in the IRIS\footnote{IRIS are improved version of IRAS maps, for
  details see \cite{Miville2005}.} maps, which appears as a slightly broken ring
with a diameter $\approx$ 1.5$^{\circ}$, and a intensity of the order of
100--200~MJy sr$^{-1}$ at 100 $\mu$m and 5--10~MJy sr$^{-1}$ at 12~$\mu$m (see
Figure~\ref{ima:iras100}).

Originally, \perseo was considered to be a supernovae remnant \citep{Pauls1989,
  Fiedler1994}, but later observations showed that this is a \ions{H}{ii} region
driven by the O9.5-B0 V star HD 278942, located at the center of the ring
\citep{Zeeuw1999}. More recent studies \citep{Andersson2000,Ridge2006})
concluded that \perseo was indeed an expanding \ions{H}{ii} bubble that has
emerged from the outer edge of the cloud.

%
\citet[][hereafter W05]{Watson2005} carried out a detailed study of the spectral
energy distribution (SED) of \perseo in a wide frequency range, from 408~MHz to
3000~GHz. This study combined, among others, the observations performed with the
COSMOSOMAS experiment at 11, 13, 15 and 17~GHz
\citep{Gallegos2001,Fernandez2006}, and the 3rd year WMAP data
\citep{Bennett2003}.  The total emitting region in \perseo was slightly resolved
in COSMOSOMAS and WMAP data, and was modelled using an ellipse with $1.6^\circ
\times 1.0^\circ$ and P.A.$=51^\circ$, centered on R.A.$=55.4^\circ$ and
Dec.$=+31.8^\circ$.
The SED for that region showed clear evidence for anomalous microwave emission,
with a peaked spectrum around 20-30~GHz, indicative of spinning dust. Indeed,
W05 showed that an adequate fit to the SED can be achieved from 10 to 50~GHz
only when including three components in the analysis: (a) optically thin
free-free emission; (b) vibrational dust emission with $T_{\rm dust} = 19$~K and
emissivity index of 1.55; and (c) a combination of the spinning dust models of
\cite{DraineED} for warm neutral medium and molecular cloud ($0.8$WNM +
$0.3$MC). No bright unresolved source that could be ultra-compact \ions{H}{ii}
region or gigahertz-peaked source could be found.

%
Detailed observations of \perseo with the Very Small Array (VSA) interferometer
at 33~GHz \citep{Tibbs2010} and an angular resolution of 10-40~arcmin, showed
that the region consists of five distinct components, all of which are found to
exhibit a emission at 33~GHz which is highly correlated with the far-infrared
emission. The most interesting result is that the VSA resolved out most of the
emission in the region, as those five components contribute to only $\approx 10$
per cent to the total flux density of the diffuse extended emission detected in
W05. Therefore, the bulk of the anomalous emission in \perseo is diffuse.

%
Concerning the polarization level of the anomalous microwave emission in the
region, the only measurement at the relevant frequencies (10-30~GHz) done so far
was presented in \cite{Battistelli2006}. Using dual orthogonal polarizations
with the COSMOSOMAS experiment, the resulting total polarization level in
\perseo at 11~GHz was found to be $\Pi = 3.4^{+1.5}_{-1.9} \%$ (95\% confidence
level). Based on this value, they concluded that this weak detection of
polarization would be associated to the spinning dust grains.
%
%
Recently, \cite{2009IAUS..259..603R} suggested that \perseo is acting as a
Faraday screen hosting a strong regular magnetic field, which would rotate the
background polarized emission. They suggest that a Faraday screen model with a
rotation measure of $RM = 190$~rad~m$^{-2}$ would explain the 11~GHz
observations, as well as their new 11~cm Effelsberg observations in the region
(see Fig.~6 in that paper).
In any case, even if the \cite{Battistelli2006} result is only an upper limit to
the polarization level of the anomalous emission, its amplitude favours the
electric dipole emission model with resonance paramagnetic relaxation (see
\cite{Lazarian2000}).

In this context, the seven-year WMAP data represent an opportunity to update the
measurements of intensity signal in the Perseus region, as well as to constrain
the polarization level of the anomalous microwave emission in the frequency
range where this emission is dominant (20-30~GHz).  As shown by
\cite{Tibbs2010}, 90\% of the emission in the region is diffuse, so the angular
resolution of WMAP data is sufficient to provide a reliable measurement.

\begin{figure*}
\centering
\includegraphics[width=5.5cm]{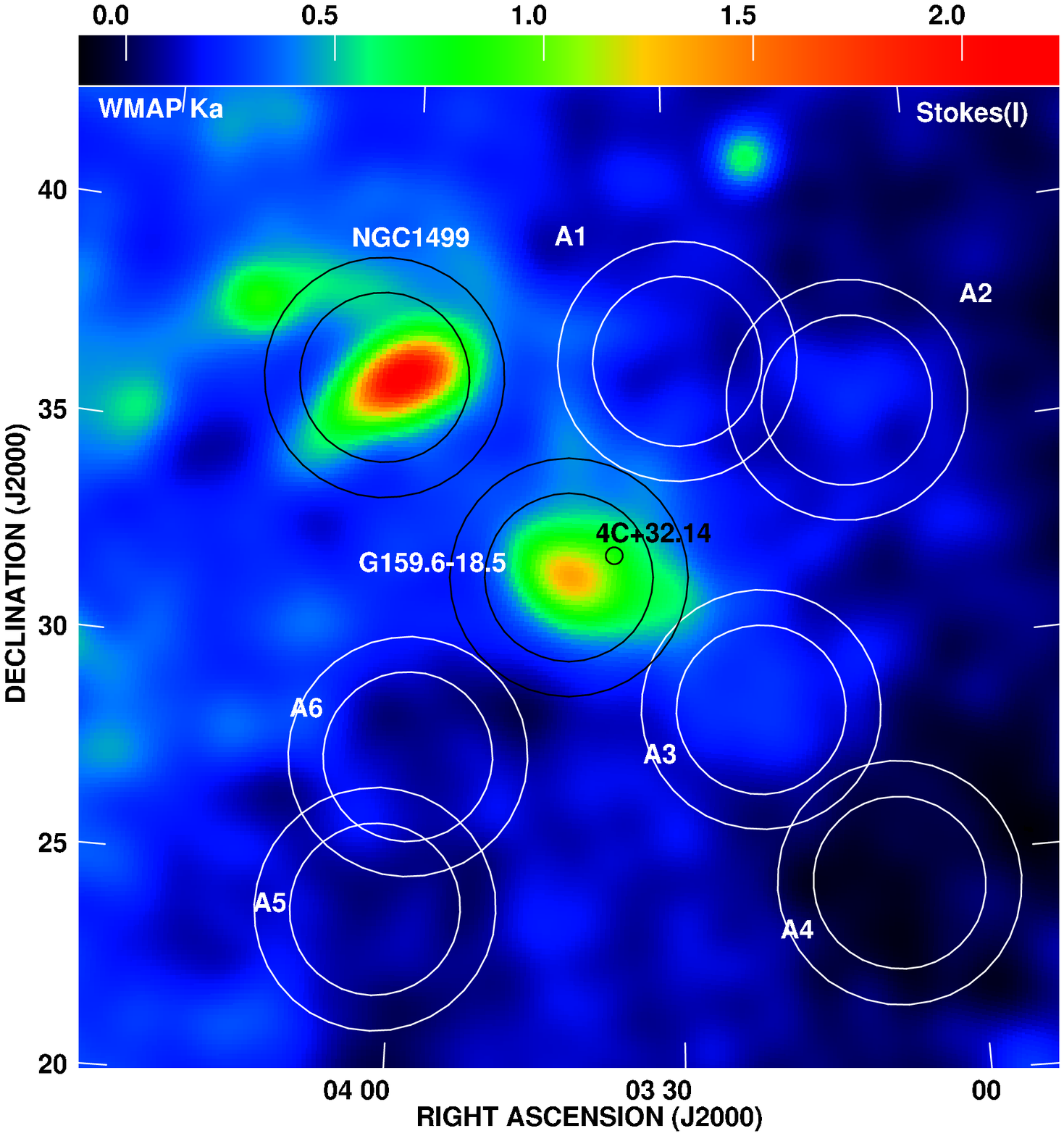}%
\includegraphics[width=5.5cm]{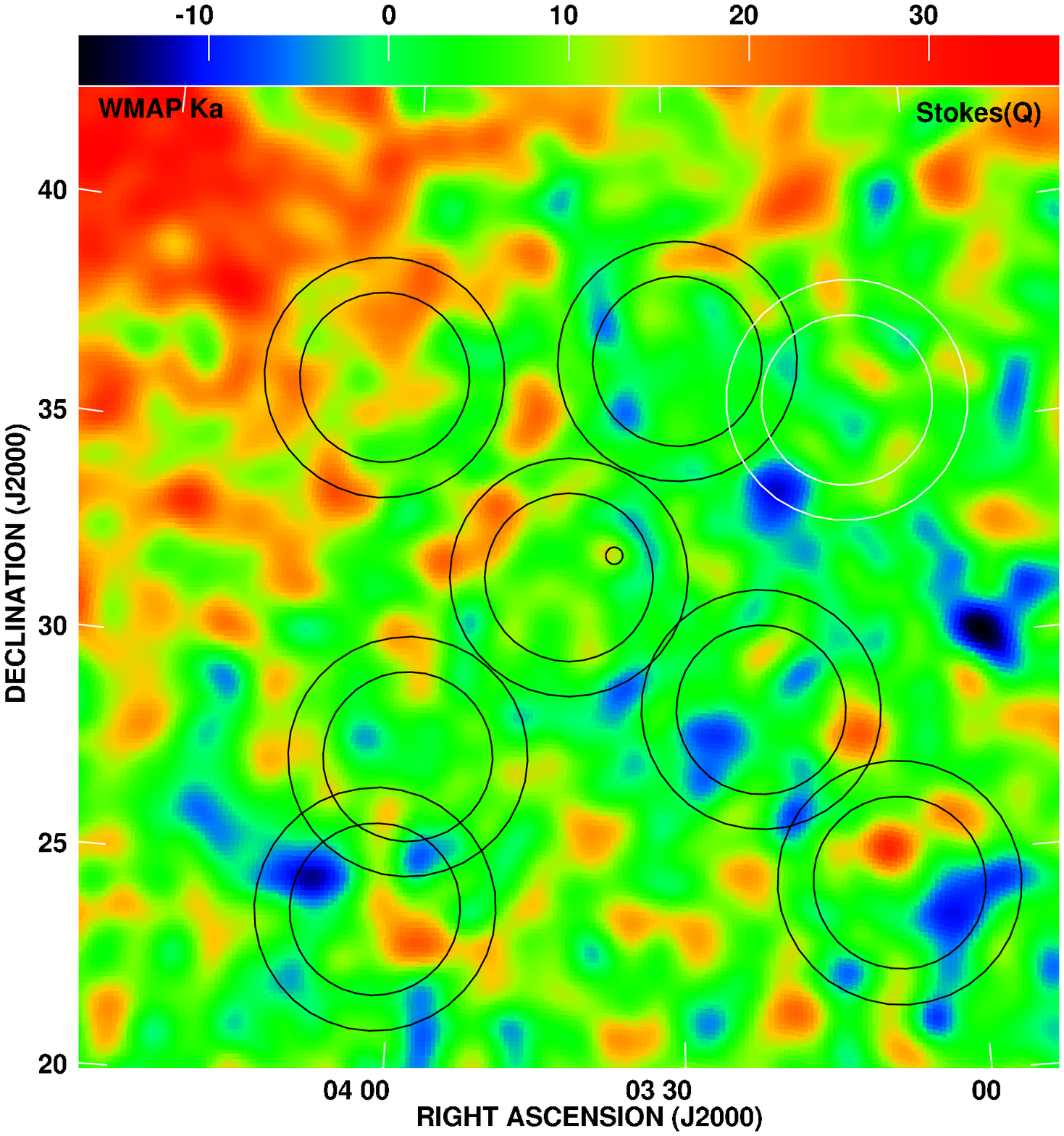}%
\includegraphics[width=5.5cm]{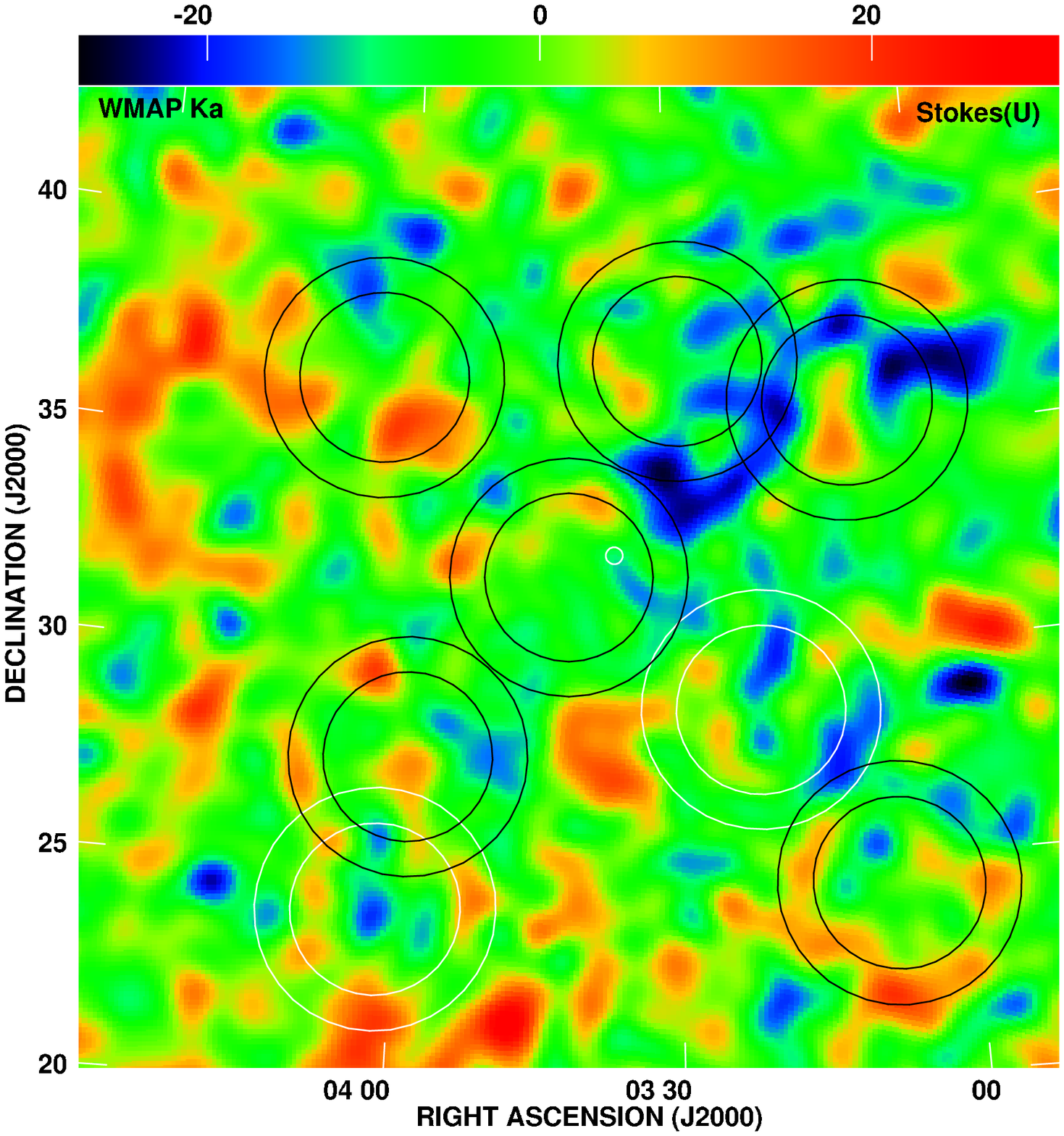}
\caption{Neighborhood of the Perseus Molecular Complex as seen by WMAP-7 at
  33~GHz (Ka band), with field of view of $23^{\circ} \times 23^{\circ}$
  centered in the \perseo region. All images (I, right panel; Q, center; and U,
  left panel) are smoothed to $1^{\circ}$ resolution, and in units of mK
  (thermodynamic). The black circles define the circular apertures around the
  objects studied in this paper (see text for details).}
\label{ima:perseocalifornia}
\end{figure*}

\section{Data and Methodology}

\subsection{WMAP data}
\label{sec:wmap7}
In this paper, we use the seven-year WMAP data products \citep{Jarosik2010},
which are available in the LAMBDA\footnote{http://lambda.gsfc.nasa.gov/} website
in the {\sc HEALPix}\footnote{http://healpix.jpl.nasa.gov} pixelisation scheme
\citep{Healpix}. In particular, we use the smoothed I, Q and U maps, for each of
the five frequency bands centered at 23, 33, 41, 61 and 94~GHz (K, Ka, Q, V and
W respectively).  The original angular resolution of these maps is approximately
$0.82$, $0.62$, $0.49$, $0.33$, and $0.21\deg$ respectively, but they are
degraded to a common resolution of $1\deg$.
For each map, a full description of the noise covariances (II, QQ, UU, QU) is
provided. This information will be used in our analyses.
We would like to stress that the WMAP polarization maps are derived from the
difference of two differential measurements. Thus, the WMAP measures a double
difference in polarized intensity, not the intensity of the difference of the
electric field as with interferometers and correlation receivers \citep[for
further details, see][]{Page2007, Jarosik2007, Kogut2003}.

From those Q and U maps at each frequency band, we can obtain the maps of the
polarization intensity as $P = \sqrt{ Q^2 \ + U^2}$, while the angle between the
polarization direction of the electric field and the Galactic meridian can be
obtained as $\gamma = \frac{1}{2} \arctan(U/Q)$.

Figure~\ref{ima:perseocalifornia} shows a detailed view of the Perseus Molecular
Complex as seen in the Ka-band of WMAP (33~GHz). In this map, two strong sources
are clearly visible: \perseo (placed at the center of the image), and the
\ions{H}{ii} region NGC1499.
This later object will be used as null hypothesis in our study, as free-free
emission is known to be unpolarized. This null-test was also used in the study
of \cite{Battistelli2006}.
Finally, the figure also displays six nearby regions which have been selected
due to their low emission in intensity. These regions will be used below to
characterize the behaviour of the background diffuse emission in the
surroundings of \perseo.

\begin{figure*}
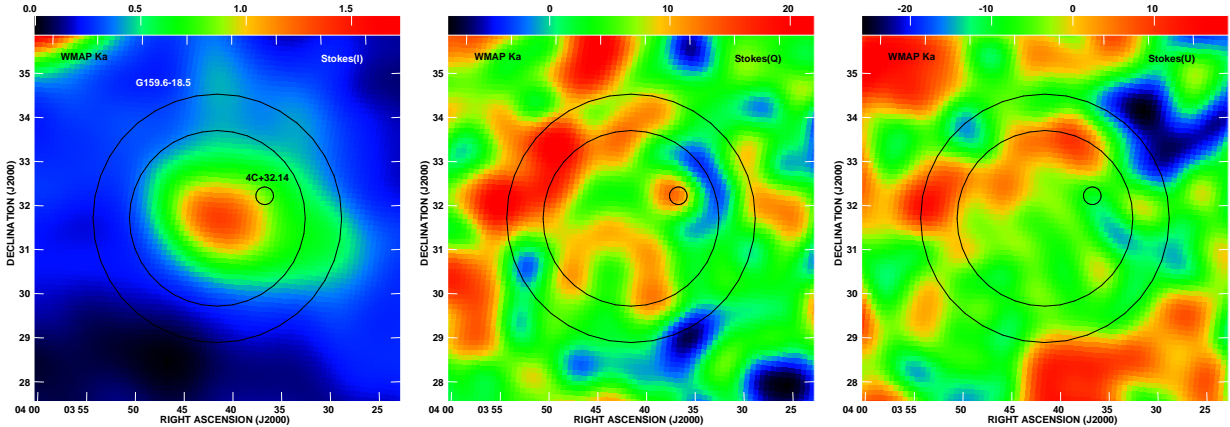

\centering
\includegraphics[width=5.5cm]{Perseo_Ka_I.ps}%
\includegraphics[width=5.5cm]{Perseo_Ka_Q.ps}%
\includegraphics[width=5.5cm]{Perseo_Ka_U.ps}
\caption{Close view of \perseo as seen by WMAP-7 at 33~GHz (Ka band), with a
  field of view of approximately $8^{\circ} \times 8^{\circ}$. The images show I
  (right frame), Q (central frame) and U (left frame) at $1^{\circ}$ resolution,
  and in units of mK (thermodynamic). The small aperture at the northwest of the
  center of \perseo corresponds to the solid angle defining the quasar
  4C+32.14. The pixels within this small aperture were not considered in our
  flux estimation.}
\label{ima:perseo}
\end{figure*}

\subsection{Ancillary data}
\label{sec:ancillary}

\subsubsection{Intensity}
As a reference for the SED in \perseo, in this work we use the data points from
Table~1 in W05. 
In that table, in addition to the COSMOSOMAS and WMAP measurements, there are
also data points at 0.408~MHz \citep{Haslam1982}; at 1420~MHz \citep{Reich1988};
and at 100, 140 and 240~$\mu$m, based on the DIRBE maps \citep{Silverberg1993}.

Here we have re-evaluated the intensity measurements on the Haslam, WMAP and
DIRBE maps, using the methodology described in Section~\ref{sec:measuring}.
The derived fluxes are slightly higher than those presented in W05 (by a factor
of 5-13 per cent), although as pointed out below, this is probably due to the
different methodology used for the flux determination in W05.

After W05, \cite{Semenova2009} used the RATAN-600 telescope to obtain
independent measurements in the region. Although an estimate of the error bar of
these new measurements is not provided in their paper, the overall shape of the
SED in \perseo in the frequency range 1-20~GHz is fully compatible with the
results of W05, showing a rising spectrum in that frequency range.

Recently, \cite{2009IAUS..259..603R} presented preliminary results of the
emission in \perseo at 11~cm and 6~cm with the Effelsberg and the Urumqi
telescopes, respectively. These results support that the emission in the
frequency range 2.7-5~GHz is compatible with an optically thin thermal gas, as
expected from the SED in W05.

\subsubsection{Polarization}
The only reported measurement to date of the radio polarized intensity in
\perseo comes from the COSMOSOMAS experiment at 11~GHz
\citep{Battistelli2006}. The measured fractional polarization is $\Pi =
3.4^{+1.5}_{-1.9} \%$ (95\% C.L.).

Recently, \cite{2009IAUS..259..603R} presented preliminary results of the
polarized emission in \perseo at 11~cm ($\sim 2.7$~GHz) with the Effelsberg
telescope. They reported a significant degree of polarization along the ring,
which they interpreted as an indication that \perseo acts as a Faraday screen
rotating the polarization angle of the background emission in the region.
However, this effect will be strongly suppressed at the peak frequencies
of the anomalous emission (20-30~GHz) as discussed below.

\cite{Kogut2007} presented a full-sky model of the polarized Galactic microwave
emission, considering in their analysis the polarization from the thermal dust
and synchrotron emissions. The resulting polarized dust model obtained in this
paper shows that the Perseus molecular complex lies near to the boundaries of
the P06 mask \citep[see mask definition in][]{Page2007}, where the expected dust
fractional polarization ($\Pi_{dust}$) reaches $6.4\pm 1.3$\% and $2.0\pm 1.4$\%
for $|b|<10\deg$ and $|b|>10\deg$, respectively, inside P06; and $3.6\pm 1.1$\%
outside the P06 mask.
This fractional degree of polarization will be of importance when studying the
high frequency behaviour of the polarized emission in the \perseo region.

\subsection{Determination of the Spectral Energy Distribution in intensity and
  polarization}
\label{sec:measuring}

Figures~\ref{ima:perseocalifornia} and \ref{ima:perseo} show that there is no
obvious structure seen within the \perseo region in the polarization maps. In
this case, we can not apply a similar method to the one used in W05 for the flux
determination. In that paper, a simple elliptical Gaussian model was used to
describe the region, and fluxes were obtained using a direct fitting to the
data.

Here, we will adopt a direct aperture integration, or ``ring analysis'', both
for the intensity and polarization determinations. The practical implementation
of this method has been previously used in other works
\citep{Bennett1993,Banday1996,HMRM04}. The basic idea is as follows. For a
certain frequency map, we define a circular aperture around the object with
radius $r_1$, defining a solid angle $\Omega_{\rm S}$ around the source. In
order to correct for the background emission, a circular corona with radius $r_1
< r < r_2$ is used. In order to preserve a similar area both for the source and
the background flux estimation, we use the relation $r_2 \ = r_1 \sqrt{2}$
between the two apertures.
This method provides an efficient correction of the background emission around
the source for apertures ($r_1$ values) comparable to the beam size, although
for larger apertures might be biased due to the background or CMB contributions.
The error bar associated to this measurement contains the quadratic sum of two
contributions, the pure instrumental noise and the increase of the variance due
to the CMB fluctuations. 
It is important to note that the evaluation of both contributions should account
for correlation terms. In the case of the CMB, this is introduced by means of
the two-point correlation function, which is evaluated using the (measured) WMAP
angular power spectrum. In the case of the noise, the correlation between pixels
appears after the smoothing process, and is taken into account in a
pixel-by-pixel basis.

As we will see below, in the case of the selected aperture for \perseo ($r_1 =
2^\circ$) and the instrumental noise level of the WMAP data, the CMB
contribution to the error bar is dominant for the intensity Stokes parameter,
while its contribution is negligible for Stokes Q and U.
Although the main focus of this paper is the polarization determination, for
illustration we have also considered how the intensity measurements change with
our aperture method when a CMB correction is performed before the flux
determination. For that discussion, we have used the Internal Linear Combination
(ILC) map provided by the WMAP team in the LAMBDA webpage \citep{Gold2010}.
Although this map does not provide an optimal separation of CMB component, it
provides a reasonable approximation which can be used to estimate its impact on
the final intensity measurements.

\section{Results and Discussion}
\label{sec:results}

\subsection{Control regions}

\subsubsection{Characterization of the background emission}
\perseo is relatively close to the galactic plane, in a region with a
non-negligible diffuse emission which varies markedly with the position.
In order to characterize the behaviour of this background emission, 
we have selected six regions with low emission in
intensity in the surroundings of the Perseus molecular complex.
The central coordinates of the six regions are shown in table~\ref{tab:coord_r}.
Around each of them, we have chosen the same aperture as for \perseo
($r_1=2^{\circ}$) for the flux determination.

\begin{table}
\centering
\caption{Identification number and coordinates of the six control regions 
  in the surroundings of \perseo used for the background characterization. 
  The aperture used in all cases and at all frequencies is $r_1 = 2^{\circ}$.}
\scriptsize{
\begin{tabular}{c c c c c c c }
ID & A1 & A2 & A3 & A4 & A5 & A6 \\
\hline
\hline
R.A. [$^{\circ}$] & 52.19 & 47.30 & 50.21 & 46.88 & 60.39 & 59.70 \\
DEC. [$^{\circ}$]& 36.88 & 35.74 & 28.56 & 24.34 & 23.88 & 27.48 \\
\hline
\end{tabular}
\label{tab:coord_r}
}
\end{table}

Table~\ref{tab:regions} summarizes our results. In addition to the I, Q, and U
determinations at each frequency, we also include in the last two columns the
values for the fluxes inside the inner aperture ($r_1$) with no background
correction. These numbers, quoted as Q$_{back}$ and U$_{back}$, will be used
below for the analysis of the Faraday rotation in the region
(Sect.~\ref{sec:result_perseo}).
The table also includes an estimation of the total linear polarization ${\rm
  P}=\sqrt{{\rm Q}^2+{\rm U}^2}$, and also the upper limit on this linear
polarization ${\rm P}_0$ that we would obtain in the null-hypothesis case using
a zero signal (${\rm Q} ={\rm U = 0}$) and the same noise levels.
As in all these values no detection is found, we quote upper limits which
correspond to the 95\% confidence level derived from a maximum likelihood
approach based on the Q and U measurements.

We first note that at all frequencies, the main contribution to the error budget
in intensity comes from the CMB part, which for illustration purposes is written
separately in the table~\ref{tab:regions} (values in brackets in the second
column). As one would expect, its relative contribution rises with approximately
$\nu^2$ dependence, as the temperature sensitivity of WMAP is comparable in all
channels.
In the case of Q and U, the CMB contribution is negligible and it is not
included in the error bar.

Finally, as a summary of the results in table~\ref{tab:regions}, we present in
table~\ref{tab:statistic} the peak-to-peak variations (I$_{p-p}$), means and the
root-mean-squares of these values for the six regions.
We note that in the case of intensity, the variations in the three first
frequencies (23, 32 and 41~GHz) is even larger than the expected CMB
contribution, which implies that in addition there is a significant contribution
to the error budget coming from the variations of the diffuse background in the
surroundings.
This effect is not seen in polarization, so in this case the error bars seem to
be dominated by the instrumental noise contribution.
However, we note that the average values of Q$_{back}$ and U$_{back}$ show a
small degree of diffuse polarization in the background, which is very well
corrected by the ring analysis.

\begin{table*}
  \centering
  \caption{Results obtained for the six control regions in the surroundings of
    G159.6-18.5. Stokes I, Q, U are measured using the ``ring analysis'' method
    with an aperture of $r_1 = 2^{\circ}$ at all frequencies. Error bars in all
    cases include the instrumental noise contribution. In the case of Stokes I,
    the number in brackets indicates the expected CMB contribution to the error
    budget, which has to be added in quadrature to the instrumental noise.
    Columns 5 and 6 present the derived upper limits (at the 95\% confidence level ) 
    on the linear polarization P, as well as upper limit P$_0$ that would be
    obtained in the null-case (i.e. using Q$=$U$=0$). Last two columns
    (Q$_{back}$ and U$_{back}$) show the direct flux determination within the
    aperture $r_1$ without the diffuse background correction (i.e. without the
    subtraction of the flux in the circular corona around $r_1$).  }
  \resizebox{0.89\textwidth}{!}{
\begin{tabular}{c c c c c c c c}
Region A1 & & & & & & &\\
$\nu$ [GHz]& I [Jy] & Q [Jy]& U [Jy]& P [Jy] & P$_0$ [Jy] & Q$_{back}$ [Jy] & U$_{back}$ [Jy]\\
\hline
\hline
23&  -3.57$\ \ \pm$   0.13 ($\pm$   1.49) &  -0.27$\pm$   0.16&  -0.27$\pm$   0.21& $<$   0.72& $<$   0.46&   1.12$\pm$   0.13&  -1.28$\pm$   0.16\\
33&  -6.05$\ \ \pm$   0.23 ($\pm$   3.00) &  -0.39$\pm$   0.30&   0.43$\pm$   0.38& $<$   1.48& $<$   0.82&   0.39$\pm$   0.26&  -0.85$\pm$   0.33\\
41&  -7.71$\ \ \pm$   0.32 ($\pm$   4.59) &  -0.02$\pm$   0.41&  -0.13$\pm$   0.54& $<$   2.10& $<$   1.17&   0.02$\pm$   0.37&  -0.86$\pm$   0.49\\
61& -13.86$\ \ \pm$   0.80 ($\pm$   9.67) &   0.09$\pm$   1.02&   0.16$\pm$   1.31& $<$   3.29& $<$   2.86&   0.93$\pm$   0.95&  -0.97$\pm$   1.22\\
94& -20.81$\ \ \pm$   1.92 ($\pm$  19.98) &  -3.98$\pm$   2.44&   0.95$\pm$   3.19& $<$   7.26& $<$   6.95&  -1.00$\pm$   2.29&   9.74$\pm$   3.00\\
\hline
 & & & & & & &\\
Region A2 & & & & & & &\\
\hline
\hline
23&   3.97$\ \ \pm$   0.13 ($\pm$   1.50) &   0.23$\pm$   0.17&   0.26$\pm$   0.21& $<$   0.70& $<$   0.46&   1.75$\pm$   0.13&  -1.30$\pm$   0.17\\
33&   5.91$\ \ \pm$   0.23 ($\pm$   3.02) &   0.01$\pm$   0.30&   0.04$\pm$   0.38& $<$   1.09& $<$   0.84&   0.68$\pm$   0.26&  -1.11$\pm$   0.33\\
41&   7.43$\ \ \pm$   0.34 ($\pm$   4.61) &  -0.26$\pm$   0.44&   0.09$\pm$   0.56& $<$   1.37& $<$   1.23&   0.36$\pm$   0.40&  -1.33$\pm$   0.51\\
61&  13.58$\ \ \pm$   0.81 ($\pm$   9.72) &  -0.78$\pm$   1.03&  -0.81$\pm$   1.35& $<$   3.33& $<$   2.93&   0.78$\pm$   0.95&  -0.10$\pm$   1.25\\
94&  28.66$\ \ \pm$   1.98 ($\pm$  20.09) &   0.40$\pm$   2.50&  -4.84$\pm$   3.29& $<$   7.92& $<$   7.16&   5.48$\pm$   2.34&   2.54$\pm$   3.10\\
\hline
 & & & & & & &\\
Region A3 & & & & & & &\\
\hline
\hline
23&   6.14$\ \ \pm$   0.13 ($\pm$   1.49) &  -0.25$\pm$   0.17&  -0.60$\pm$   0.23& $<$   1.02& $<$   0.49&   1.03$\pm$   0.13&  -1.74$\pm$   0.18\\
33&   5.59$\ \ \pm$   0.24 ($\pm$   2.99) &  -0.20$\pm$   0.31&   0.47$\pm$   0.41& $<$   1.35& $<$   0.89&   0.32$\pm$   0.27&  -0.52$\pm$   0.35\\
41&   6.40$\ \ \pm$   0.36 ($\pm$   4.57) &  -0.06$\pm$   0.46&  -0.07$\pm$   0.59& $<$   1.79& $<$   1.29&   0.42$\pm$   0.41&  -1.07$\pm$   0.54\\
61&   8.86$\ \ \pm$   0.89 ($\pm$   9.63) &   0.48$\pm$   1.13&   1.22$\pm$   1.48& $<$   4.54& $<$   3.21&   1.31$\pm$   1.05&   0.52$\pm$   1.39\\
94&  14.65$\ \ \pm$   2.17 ($\pm$  19.91) &   0.41$\pm$   2.75&   0.22$\pm$   3.59& $<$   9.90& $<$   7.81&  -1.84$\pm$   2.59&   5.91$\pm$   3.39\\
\hline
 & & & & & & &\\
Region A4 & & & & & & &\\
\hline
\hline
23&  -2.90$\ \ \pm$   0.14 ($\pm$   1.49) &   0.00$\pm$   0.18&  -0.74$\pm$   0.24& $<$   1.14& $<$   0.51&   1.14$\pm$   0.14&  -1.51$\pm$   0.19\\
33&  -3.41$\ \ \pm$   0.26 ($\pm$   2.99) &   0.18$\pm$   0.33&  -0.36$\pm$   0.43& $<$   1.54& $<$   0.93&   0.72$\pm$   0.28&  -0.17$\pm$   0.37\\
41&  -4.45$\ \ \pm$   0.41 ($\pm$   4.58) &   0.74$\pm$   0.50&   1.35$\pm$   0.75& $<$   2.40& $<$   1.58&   1.09$\pm$   0.46&   0.79$\pm$   0.68\\
61&  -7.20$\ \ \pm$   0.91 ($\pm$   9.65) &  -0.03$\pm$   1.15&  -2.88$\pm$   1.51& $<$   4.79& $<$   3.28&   0.93$\pm$   1.07&  -2.89$\pm$   1.40\\
94& -11.37$\ \ \pm$   2.26 ($\pm$  19.94) &   1.06$\pm$   2.88&   1.03$\pm$   3.71& $<$   9.35& $<$   8.12&   5.22$\pm$   2.71&  12.62$\pm$   3.48\\
\hline
 & & & & & & &\\
Region A5 & & & & & & &\\
\hline
\hline
23&  -2.02$\ \ \pm$   0.13 ($\pm$   1.50) &   0.22$\pm$   0.17&  -0.21$\pm$   0.22& $<$   0.66& $<$   0.48&   1.39$\pm$   0.14&   0.09$\pm$   0.17\\
33&  -1.65$\ \ \pm$   0.24 ($\pm$   3.02) &   0.25$\pm$   0.31&  -0.47$\pm$   0.40& $<$   0.96& $<$   0.87&   0.73$\pm$   0.27&  -0.29$\pm$   0.35\\
41&  -1.83$\ \ \pm$   0.36 ($\pm$   4.61) &   0.90$\pm$   0.48&  -0.26$\pm$   0.57& $<$   2.04& $<$   1.28&   1.07$\pm$   0.43&  -0.10$\pm$   0.51\\
61&  -2.10$\ \ \pm$   0.86 ($\pm$   9.72) &  -0.31$\pm$   1.11&   0.20$\pm$   1.38& $<$   4.95& $<$   3.05&  -0.02$\pm$   1.03&   1.20$\pm$   1.28\\
94& -11.06$\ \ \pm$   2.12 ($\pm$  20.08) &   1.81$\pm$   2.74&  -2.49$\pm$   3.38& $<$   8.07& $<$   7.50&   7.06$\pm$   2.59&   2.43$\pm$   3.19\\
\hline
 & & & & & & &\\
Region A6 & & & & & & &\\
\hline
\hline
23&  -3.95$\ \ \pm$   0.13 ($\pm$   1.49) &   0.15$\pm$   0.17&  -0.13$\pm$   0.22& $<$   0.57& $<$   0.48&   1.46$\pm$   0.13&  -0.47$\pm$   0.17\\
33&  -2.91$\ \ \pm$   0.24 ($\pm$   3.00) &  -0.05$\pm$   0.30&  -0.28$\pm$   0.40& $<$   1.17& $<$   0.86&   0.80$\pm$   0.26&  -0.39$\pm$   0.34\\
41&  -2.48$\ \ \pm$   0.35 ($\pm$   4.59) &  -0.42$\pm$   0.45&   0.94$\pm$   0.56& $<$   1.56& $<$   1.25&   0.65$\pm$   0.40&   1.16$\pm$   0.51\\
61&  -1.34$\ \ \pm$   0.83 ($\pm$   9.68) &   0.17$\pm$   1.06&  -0.09$\pm$   1.37& $<$   3.44& $<$   2.99&  -0.26$\pm$   0.98&   1.31$\pm$   1.28\\
94&  -5.80$\ \ \pm$   2.04 ($\pm$  20.00) &   1.73$\pm$   2.59&   1.37$\pm$   3.36& $<$   7.84& $<$   7.36&   4.34$\pm$   2.44&   5.86$\pm$   3.17\\
\hline
\end{tabular}
}
\label{tab:regions}
\end{table*}

\begin{table*}
  \centering
  \caption{Summary of the statistics for the six control regions in
    table~\ref{tab:regions}. We present the peak-to-peak variations (I$_{p-p}$),
    means and standard deviations of the Stokes I, Q and U, together with the
    mean of the Q$_{back}$ and U$_{back}$ values for each of the five WMAP
    frequencies. All quoted values are given in Jy. }
\resizebox{0.89\textwidth}{!}{
\begin{tabular}{c c c c c c c c c c | c c}
\hline
$\nu$ [GHz] & I$_{p-p}$ & $\langle{\rm I}\rangle$ & $\langle{\rm I}^2\rangle^{1/2}$ & Q$_{p-p}$ &$\langle{\rm Q}\rangle$ & $\langle{\rm Q}^2\rangle^{1/2}$ & U$_{p-p}$ & $\langle{\rm U}\rangle$ & $\langle{\rm U}^2\rangle^{1/2}$ & $\langle{\rm Q}_{back}\rangle$ [Jy] & $\langle{\rm U}_{back}\rangle$[Jy]\\
\hline
\hline
23&  10.10&  -0.39&   4.32&   0.50&   0.01&   0.23&   1.00&  -0.28&   0.36&    1.32&  -1.04\\
32&  11.96&  -0.42&   4.99&   0.64&  -0.03&   0.24&   0.95&  -0.03&   0.41&    0.61&  -0.55\\
41&  15.13&  -0.44&   6.06&   1.32&   0.15&   0.54&   1.61&   0.32&   0.66&    0.60&  -0.24\\
61&  27.44&  -0.34&  10.12&   1.26&  -0.06&   0.44&   4.10&  -0.37&   1.39&    0.61&  -0.15\\
93&  49.46&  -0.96&  18.70&   5.80&   0.24&   2.16&   6.21&  -0.63&   2.50&    3.21&   6.52\\
\hline
\end{tabular}
}
\label{tab:statistic}
\end{table*}

\subsubsection{NGC 1499: Null-hypothesis in polarization}
\label{sec:ngc1499}

The California Nebula is a \ions{H}{ii} region close to G159.6-18.5 (its optical
coordinates are RA(J2000)$=04^h 03^m 18^s$ and Dec$=+36^{\circ} 25.3'$). Since
the dominant emission mechanism in diffuse \ions{H}{ii} regions at these
frequencies is the free-free emission, they are expected to be practically
unpolarized, so \california will be our null-hypothesis for the polarization.

California is also an extended object at this resolution. For consistency, we
use the same aperture $r_1 = 2^{\circ}$ for all frequency bands, which after
visual inspection, is found to fit well with the solid angle that defines the
source. 
The central coordinates of the emission used for the flux integration are
RA$=60.40^\circ$ and Dec$=+36.3675^\circ$, practically coinciding with the peak
emission in the 33~GHz map.
Our results on the I, Q, U measurements, and the derived upper limits on the
linear polarization\footnote{$\Pi$ is the fractional linear polarization,
  defined as $\Pi \equiv 100\ {\rm P/I}$.} P and $\Pi$, as well as the null-case
$P_0$ and $\Pi _0$, are presented in table~\ref{tab:california}.
%

A simple extrapolation of the DIRBE fluxes to the WMAP frequency range shows
that the thermal dust emission is expected to be negligible between 20-100~GHz.
Thus, we expect that the main emission mechanism is the free-free emission.
To check this prediction, we have derived the spectral index that best-fits the
data obtained in the range 23-41~GHz, using a power law model (${\rm I} \propto
\nu^{\beta}$). We find $\beta =0.08 \pm 0.10$, which is apparently in
contradiction with the expected frequency behaviour for an optically thin
free-free emitting region at these frequencies ($\beta \approx -0.1$).
However, this seems to be a effect of the residual CMB contribution to these
measurements. If we correct the WMAP maps from the CMB emission by subtracting
the ILC map, the flux estimates at 23, 32 and 41~GHz are now found to be 61.6,
59.6 and 59.5~Jy, respectively. Using these values, we find $\beta= -0.07 \pm
0.11$, which is now compatible with the expected value.

Concerning the polarization measurements, at all frequencies the Stokes Q and U
parameters are found to be compatible with a zero level. The upper limits on P,
and $\Pi$ at the 95\% confidence level are similar to those obtained for the
null-case hypothesis (P$_{0}$ and $\Pi_0$). Indeed, the full posterior
distributions for these parameters are found to be very similar, implying a
non-detection of polarization in the region, as one would expect for an
\ions{H}{ii} region.

\begin{table*}
\centering
\caption{Summary of flux measurements for NGC1499. We use an aperture of $r_1 = 2^{\circ}$
  at all frequencies to integrate the fluxes for I, Q and U
  Stokes parameters. The upper limits on P and $\Pi$ are derived using a maximum
  likelihood analysis. Similarly, we get the upper limits P$_0$ and $\Pi_0$ for
  the null-case hypothesis (Q=U=0). As in Table~\ref{tab:regions}, for the
  intensity column the error budget is separated in two contributions, the
  instrumental noise and the CMB error (within parenthesis). }
\begin{tabular}{c c c c c c c c }
$\nu$ [GHz]& I [Jy] & Q [Jy]& U [Jy]& P [Jy] & $\Pi$ [\%] & P$_0$ [Jy] &  $\Pi_0$[\%] \\
\hline
\hline
23&  64.65$\ \ \pm$   0.13($\pm$   1.50)&   0.28$\pm$   0.16&  -0.12$\pm$   0.21& $<$   0.63& $<$   0.98& $<$   0.46& $<$   0.71\\
33&  65.70$\ \ \pm$   0.23($\pm$   3.02)&  -0.22$\pm$   0.30&   0.44$\pm$   0.37& $<$   0.90& $<$   1.40& $<$   0.82& $<$   1.27\\
41&  68.92$\ \ \pm$   0.32($\pm$   4.62)&   0.36$\pm$   0.41&   0.15$\pm$   0.54& $<$   1.18& $<$   1.84& $<$   1.17& $<$   1.84\\
61&  70.73$\ \ \pm$   0.79($\pm$   9.73)&   0.58$\pm$   1.01&   0.21$\pm$   1.31& $<$   3.05& $<$   4.73& $<$   2.84& $<$   4.43\\
94&  65.61$\ \ \pm$   1.93($\pm$  20.11)&   2.64$\pm$   2.45&  -1.55$\pm$   3.19& $<$   7.10& $<$  10.90& $<$   6.91& $<$  10.70\\
\hline
\end{tabular}
\label{tab:california}
\end{table*}

\subsection{Perseus}
\label{sec:result_perseo}

Figure~\ref{ima:perseo} shows a detailed view of the \perseo region. The
circular aperture used for the flux determination corresponds to a radius of
$r_1 = 2^{\circ}$, and is centered in RA(J2000)$=55.4^{\circ}$ and
Dec$=31.8^{\circ}$.
Since \perseo is a extended object, we have used the same aperture $r_1$ at all
frequency bands. The subset of pixels within the small aperture at the northwest
of the center of \perseo were removed for the flux determination, as they
correspond to the location of the quasar 4C+32.14. This small aperture uses
$r_{4C+32.14}=0.25^{\circ}$ at all frequencies, and it is centered at
RA(J2000)$=54.125^{\circ}$ and Dec$=32.308^{\circ}$.
Table~\ref{tab:perseo} summarizes the results obtained for the region.

\begin{table*}
\centering
\caption{ Summary of flux measurements for the \perseo region. The meaning of
  the different columns is the same as in Table~\ref{tab:california}. }
\begin{tabular}{c c c c c c c c }
$\nu$ [GHz]&I  [Jy]  & Q [Jy] & U [Jy]& P [Jy] & $\Pi$ [\%] & P$_0$ [Jy] &  $\Pi_0$[\%] \\
\hline
\hline
23&  47.76$\ \ \pm$   0.13($\pm$   1.44)&  -0.07$\pm$   0.16&   0.01$\pm$   0.22& $<$   0.48& $<$   1.01& $<$   0.47& $<$   0.98\\
33&  42.97$\ \ \pm$   0.23($\pm$   2.90)&   0.05$\pm$   0.29&   0.43$\pm$   0.39& $<$   0.86& $<$   1.79& $<$   0.85& $<$   1.76\\
41&  37.88$\ \ \pm$   0.33($\pm$   4.44)&  -0.09$\pm$   0.42&  -0.41$\pm$   0.55& $<$   1.29& $<$   2.69& $<$   1.19& $<$   2.48\\
61&  36.36$\ \ \pm$   0.80($\pm$   9.35)&   0.06$\pm$   1.01&   0.67$\pm$   1.33& $<$   3.44& $<$   7.23& $<$   2.88& $<$   6.03\\
94&  70.18$\ \ \pm$   1.94($\pm$  19.32)&   1.13$\pm$   2.46&  -2.06$\pm$   3.24& $<$   7.52& $<$  15.64& $<$   7.04& $<$  14.65\\
\hline
\end{tabular}
\label{tab:perseo}
\end{table*}

\subsubsection{Intensity Signal}

The second column in Table~\ref{tab:perseo} presents the results for the
intensity measurements in the \perseo region. A comparison with the results in
W05 shows that our fluxes are between 5 and 13 per cent higher in the first four
channels (23 to 61~GHz), and 10 per cent smaller in the last channel
(94~GHz). This small discrepancy is probably due to the different methodologies
used to derive the fluxes. In W05, an elliptical model for the region was
adopted, so any departure of the actual shape of the region from this model
might introduce these changes.

\begin{figure}
\centering
\includegraphics[width=9 cm]{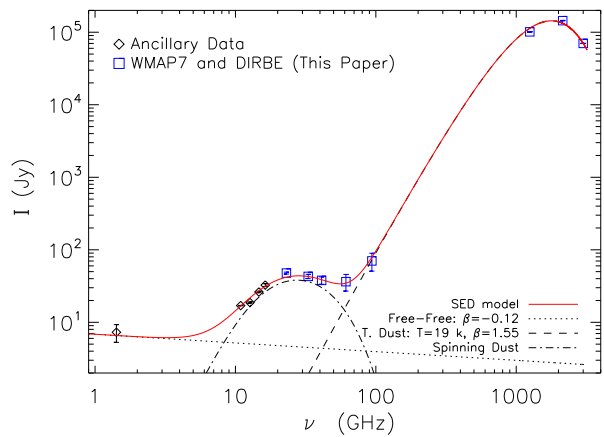}
\caption{Spectral Energy Distribution (SED) for the \perseo region. The blue
  squares and black diamond correspond to measurements obtained in this paper
  (WMAP and DIRBE bands) and the ancillary data used (COSMOSOMAS and HASLAM
  bands), respectively. The SED is fitted with three contributions: i) the
  free-free emission (${\rm I}_{ff}$), modelled as a power law with spectral
  index $\beta_{ff}=-0.12$; ii) the thermal dust emission, modelled as a
  modified black body (${\rm I}_{dust}$) with $\beta_{dust} = 1.55$ and ${\rm
    T}_{dust} = 19\ {\rm K}$; and iii) the spinning dust due to the electric
  dipole, modelled as the linear combination of two \cite{DraineED} models for
  warm neutral medium and molecular cloud ($0.85 WNM + 0.35 MC$). The red line
  represents the co-added SED model of the three contributions.}
\label{figtab:perseoi}
\end{figure}

Figure~\ref{figtab:perseoi} shows our results for the intensity signal in WMAP
and DIRBE maps, together with the ancillary data described in
section~\ref{sec:ancillary}.
Following W05, using the W band and the DIRBE data we can fit the thermal dust
emission with a modified blackbody function ${\rm I}_{dust} \propto
\nu^{\beta_{dust} + 2}\ {\rm B}(\nu,{\rm T}_{dust})$, where ${\rm T}_{dust}$ is
the dust temperature and $\beta_{dust}$ the dust emissivity index. The black
dashed line represents the best-fit model of the thermal dust emission, which
has parameters $\beta_{dust} = 1.55$ and ${\rm T}_{dust} = 19\ {\rm K}$, in full
agreement with W05. These values are expected for dust in a 'warm neutral
medium' WNM).

The low frequency behaviour is fitted by the dotted line to a free-free
emission, using the measurement at 1420~MHz ($7.3 \pm 2.0$~Jy) and the typical
free-free spectral index $\beta_{ff} = -0.12$ for an optically thin region.

As in W05, the extrapolations of the thermal dust and free-free emissions can
not explain the rising spectrum between 10 and 23~GHz, and a third component has
to bee included in the fit. Following W05, we consider a model based on a linear
combination of the \cite{DraineED} models for Warm Neutral Medium (WNM) and
Molecular Cloud (MC).
Using $\tau_{3000}=8.45 \times 10^{-4}$ (obtained from the fit thermal dust
emission) and the canonical factor of $2.13 \times 10^{24}\ {\rm H \ cm}^{-2} =
1\tau_{100}$ \citep[from ][]{Finkbeiner2004}, we obtain the column number
density of hydrogen atoms $N({\rm H})= 1.8 \times 10^{21}$~cm$^{-2}$, which
differs from the value obtained by W05 ($1.3 \times 10^{22}$~cm$^{-2}$) due to
the dependence of $N({\rm H})$ on the solid angle. 
Finally, considering our $N({\rm H})$ value and using the solid angle $\Omega_s$
that defines the source, we find that the best-fit to the observed data is
obtained with the combination ${\rm I}_{a} \approx 0.85 {\rm WNM} + 0.35 {\rm
  MC}$ (black dot-dashed line in Figure~\ref{figtab:perseoi}).
The red solid line in Fig.~\ref{figtab:perseoi} corresponds to the co-added
spectral energy distribution model ${\rm I}_{tot}$ for the three components
(${\rm I}_{tot}={\rm I}_{a} + {\rm I}_{dust} + {\rm I}_{ff}$).

\subsubsection{Polarization Signal}

Table~\ref{tab:perseo} summarizes our polarization measurements, which show that
the polarized intensity in \perseo is compatible with zero at the five frequency
bands.  Because of this reason, all quoted values in columns 5 to 8 correspond
to upper limits at the 95 per cent confidence level.
Nevertheless, these upper limits on the linear polarization fraction still
provide extremely important constraints on the nature of the anomalous emission.

\begin{figure}
\centering
\includegraphics[width=8 cm]{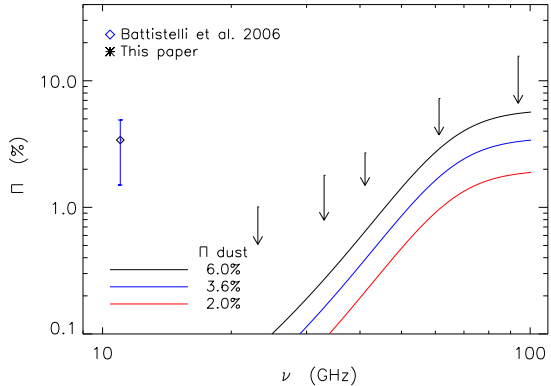}
\caption{Our measurements on the total fractional linear polarization $\Pi$
  (black arrows) in the \perseo region.  We also include the result of
  $\Pi=3.4_{-1.4}^{+1.8}$\% at 11~GHz from \cite{Battistelli2006}.  For
  illustration, the solid lines show the predicted fractional linear
  polarization due to thermal dust in the region for three values of
  $\Pi_{dust}$, namely 6.0\% (black line), 3.6\% (blue line) and 2.0\% (red
  line), which are taken to be consistent with range of values allowed by
  \cite{Kogut2007}. At the frequencies around the peak of the anomalous
  emission, the constraints are of the order of 1 per cent, and can be directly
  translated into constraints on the polarization of the anomalous emission. }
\label{figtab:perseopi}
\end{figure}

Figure~\ref{figtab:perseopi} represents our upper limits from WMAP data,
together with the 11~GHz result of \cite{Battistelli2006}. For comparison, we
also include the expected dust fractional polarization for three cases, namely
$\Pi_{dust}$ equal to 6.0\% (solid black line), 3.6\% (solid blue line) and
2.0\% (solid red line). This range of values for the polarized dust emission in
the region are taken to be consistent with the estimates in \cite{Kogut2007}.
The conclusion is that our upper limits on the fractional linear polarization in
\perseo at these microwave wavelengths can be directly translated into upper
limits on the anomalous emission, at least in the frequency range 23-41~GHz,
while the dust contribution has to be taken into account at higher
frequencies. Thus, the upper limits at 23 and 32~GHz of $\Pi(23~{\rm GHz}) <
1.0$\% and $\Pi(32~{\rm GHz}) < 1.8$\% (95\% C.L.) constitute the most stringent
constraints on the polarization of this emission.

\paragraph{Faraday screen model. } 
As mentioned in Sect.~\ref{sec:perseusregion}, \cite{2009IAUS..259..603R} have
recently suggested that \perseo is acting as a Faraday screen (FS) hosting a
strong regular magnetic field, which would rotate the background polarized
emission, being the responsible mechanism of the signal detected in
\cite{Battistelli2006}.
However, this mechanism produces negligible effects on the expected polarized
emission at the WMAP frequencies. To show this, we can use the estimates of the
background emission from the last two columns in table~\ref{tab:statistic}, and
take the rotation measure $RM=190$~rad~m$^{-2}$ as a face value to evaluate the
effect. Being conservative, we can assume that the full background emission is
rotated by the FS. In that case, for the relevant rotation angles (computed as
$\theta = RM \lambda^2$), we obtain that the expected polarization signal is
smaller than 0.1~Jy (or 0.2\%) at 23~GHz, and becomes even smaller at higher
frequencies (Ka, Q and V). Therefore, we can safely neglect its contribution at the WMAP
frequency bands.

\begin{figure}
\centering
\includegraphics[width=9cm]{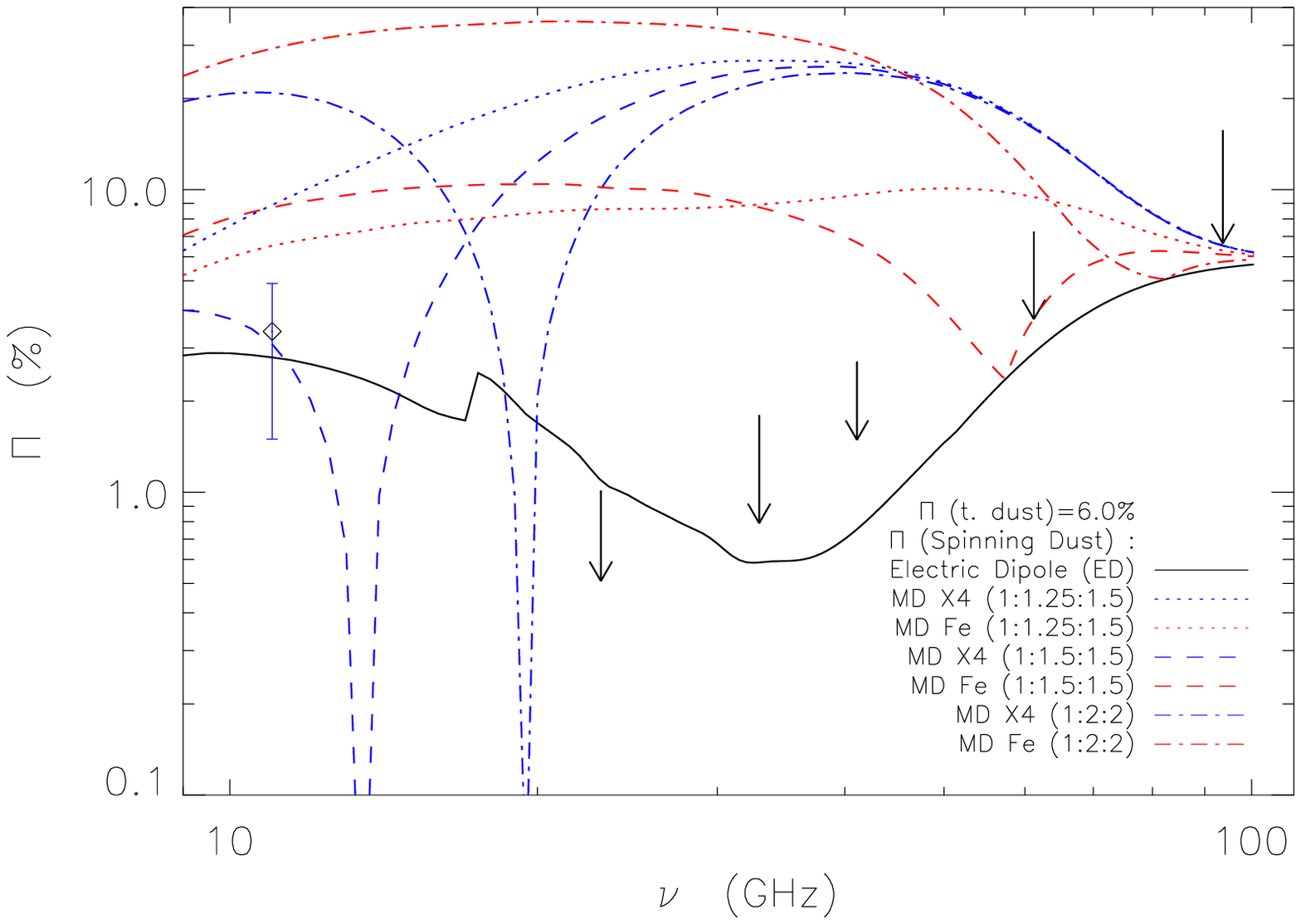}
\caption{Constraints on several anomalous emission models based on our results
  in the \perseo region.  Data points and upper limits have the same meaning as
  in Fig.~\ref{figtab:perseopi}.  The lines correspond to the fractional linear
  polarization for different spinning dust models: i) the electric dipole from
  \cite{Lazarian2000} (black solid line); and ii) the magnetic dipole emission
  from \citep{DraineMD}, that predicts a different frequency behaviour (blue and
  red lines) depending on the composition and shape of the grains.  For
  completeness, we have added to all models the contribution of polarized
  thermal dust emission with a polarization fraction of $\Pi_{dust}=6.0$\%.  Our
  results exclude the magnetic dipole emission models as the physical process
  responsible of the observed polarization in \perseo. In contrast, these are
  consistent with the expected linear polarization from the electric dipole
  emission and resonance relaxation.  }
\label{figtab:perseopi2}
\end{figure}

\paragraph{Implications on the anomalous emission models. }
Figure~\ref{figtab:perseopi2} illustrates the implications of our measurements
on the existing models of anomalous emission.
The figure also includes the value of \cite{Battistelli2006}, because even in
the case that this result only provide an upper limit to the polarization of the
anomalous emission, it still is a very valuable constraint.

In Figure~\ref{figtab:perseopi2} we have considered two families of the
so-called {\it spinning dust} models: i) the electric dipole emission and the
resonance relaxation (hereafter ED) proposed by \cite{Lazarian2000}; and ii) the
magnetic dipole emission (hereafter MD) proposed in \cite{DraineMD}. For this
second case, the number of variants is very large, so here we will restrict
ourselves to those models where the grains that dominate the 10-100~GHz emission
consist of a single magnetic domain. These models were computed by
\cite{DraineMD} for metallic Fe, and for an hypothetical material X4 (defined by
them). Both results will be used here, for three different grain shapes, namely
$1:1.25:1.5$ (dotted lines), $1:1.5:1.5$ (dashed lines) and $1:2:2$
(dot-and-dashed lines). For further details on these models, see
\cite{DraineMD}.

We note that in addition to the anomalous emission, when plotting the different
models in figure~\ref{figtab:perseopi2} we have also included the contribution
of the polarized thermal dust emission, using the value $\Pi_{dust}=6.0$\%. In
practise, this is done as described by this equation
\begin{equation}
 \Pi_{m} =\frac{{\rm P}_{a}+{\rm P}_{dust}}{{\rm I}_{a} + {\rm I}_{dust} + {\rm I}_{ff}}
\label{eq:fptotal}
\end{equation}
where the model for the linear polarization of anomalous and the thermal dust
emission can be represented by ${\rm P}_{a} = {\rm I}_a \Pi_a$ and ${\rm
  P}_{dust} = {\rm I}_{dust} \Pi_{dust}$ respectively.  The ${\rm I}_{dust}$,
I$_{ff}$ and ${\rm I}_{dust}$ contributions were obtained in the previous
section.

\begin{figure}
  \centering
  \includegraphics[width=9cm]{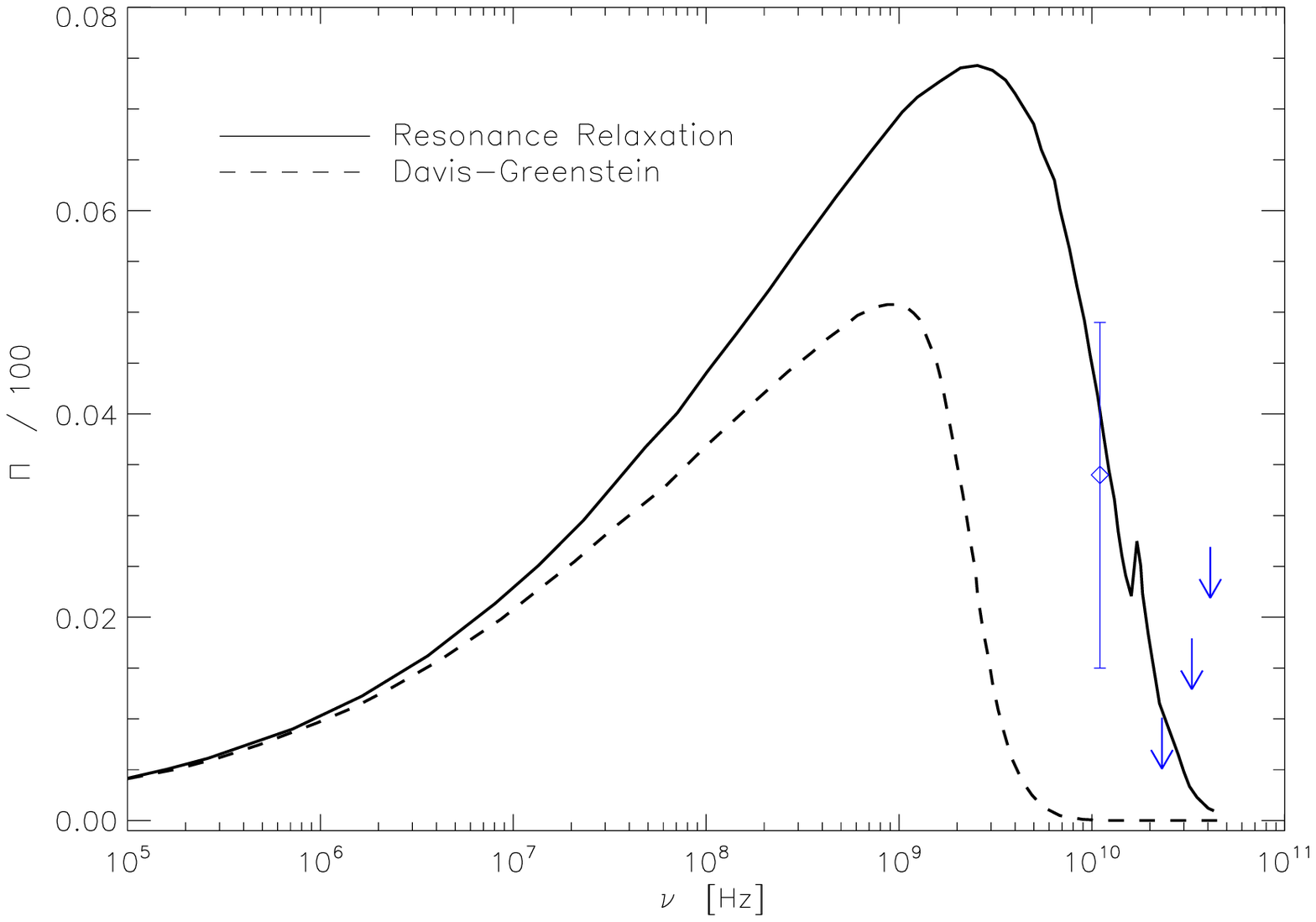}
  \caption{Constraints on the grain alignment for both resonance and
    Davis-Greenstein relaxation models for grains in the cold interstellar
    medium as a function of the frequency.  Both theoretical curves are taken
    from \cite{Lazarian2000}. For the resonance relaxation the saturation
    effects are neglected, which means that the upper curves correspond to the
    maximal values allowed by the paramagnetic mechanism. The data point
    corresponds to the \cite{Battistelli2006} result, while the three upper
    limits are the results in this paper. }
\label{figtab:resonancedavis}
\end{figure}

From Figure~\ref{figtab:perseopi2} we see that the measurement at 11~GHz can not
completely rule out the model of dust grains of X4 with semiaxes
1:1.5:1.5. However, when including the upper limits derived here from WMAP data,
all the considered models based on magnetic dipole emission can be ruled out.
In contrast, all these measurements are in agreement with the predictions of the
polarization fraction from the electric dipole and resonance relaxation theory
\citep{Lazarian2000} at this frequency range (see
Fig.~\ref{figtab:resonancedavis}).

\section{Conclusions}
A detailed understanding of the physical mechanism responsible for the anomalous
microwave emission is still needed. The polarization properties of those regions
with anomalous emission, as \perseo, are very useful tools to disentangle among
the different models proposed in the literature.

Here, we used the WMAP7 data to study the intensity and polarization properties
of the emission in the \perseo region. 
In intensity, our results confirm the presence of the anomalous microwave
emission. We present an updated SED model that takes into account the
contribution of the electric dipole emission of very small dust grains rapidly
rotating \citep{DraineED}.

Concerning the polarization in the region, we present the first constraints on
the polarization properties of the anomalous microwave emission at high
frequencies (23-94~GHz).
Due to the fact that more of the 90 per cent of the emission in \perseo is
diffuse and extended over scales larger than 40~arcmin \citep{Tibbs2010}, the
angular resolution of WMAP is not a limitation when constraining the
polarization degree of the emission in the region.
Although we find no detection of polarization in any of the five frequency
bands, the derived upper limits allow to exclude a significant number of models
based on the magnetic dipole emission of dust grains \citep{DraineMD} as the
physical process responsible of the observed polarization.

The main conclusion is that the combination of our constraints in the range
23-94~GHz, with the the measurement at 11~GHz from the COSMOSOMAS experiment
\citep{Battistelli2006} are consistent with the expected linear polarization
arising from the electric dipole model with resonance relaxation.

Further observations with higher sensitivity and angular resolution in this and
other regions will be valuable to understand the nature of this anomalous
emission in polarization.

\acknowledgments 

We acknowledge the use of the Legacy Archive for Microwave Background Data
Analysis (LAMBDA). Support for LAMBDA is provided by the NASA Office of Space
Science.
Some of the results in this paper have been derived using the {\sc HEALPix}
\citep{Healpix} package.
This work has been partially funded by project AYA2007-68058-C03-01 of the
Spanish Ministry of Science and Innovation (MICINN).  JAR-M is a Ram\'on y Cajal
fellow of the Spanish Ministry of Science and Innovation (MICINN).

\bibliographystyle{apj}

\end{document}